\documentstyle[12pt]{article}
\begin{document}
\input feynman
\begin{center}
\vskip 1.5 truecm
{\bf SOLUTION OF TWIST-3 EVOLUTION EQUATION IN DOUBLE LOGARITHMIC
APPROXIMATION IN QCD }\\
\vspace{.5cm}
A.G.Shuvaev \\
\vspace{.5cm}
Petersburg Nuclear Physics Institute, \\
Gatchina, St.Petersburg 188300 Russia \\
\end{center}
\vspace{1cm}
\begin{abstract}
The solution of DGLAP evolution equation for the twist-3 gluon operators
is obtained in the Double Logarithmic Approximation of QCD perturbation
theory. The method used for the solution is similar to the reggeon field
theory. The asymptotics of the twist-3 parton correlation function for
the small Bjorken variables $x_B$ is found.
\end{abstract}

\noindent
1. The growth of the structure function $f(x_B,Q^2)$ in the region of small
Bjorken variables $x_B$ makes it necessary to take into account the high
twist contribution. According to the operator product expansion the local
operator of spin $J$ contributes to the number $J$ Mellin moment of
the structure function $f(x_B,Q^2)$. The $Q^2$-dependence of the operators
is perturbatively determined by evolution equation collecting in the leading
order (LLA) the terms of the form $(\alpha_S\ln Q^2/\mu^2)^n$.
The asymptotic behavior for $x_B \to 0$ is governed by the rightmost
singularity of the anomalous dimension in the variable $J$ continued to
the complex plane. However the calculation of the anomalous dimension
becomes more complicated for twists $N \ge 3$ because of the large number
of the local operators.

There is another approach proposed in Ref.\cite{Jar} and based on
the relation of small-$x_B$ behavior of the structure function and BFKL
equation summing up the powers $(\alpha_S\ln 1/x_B)^n$. It enables to find
twist-2 anomalous dimension near the singularity position $J \to 1$
for the leading and next to leadings orders in terms of $\ln Q^2$.
The situation is also more involved for the higher twists $N \ge 3$ since one
has to solve the equation for $N$ reggeized gluons \cite{BJKP} and then
to extract from it the anomalous dimension.

\noindent
2. Here we consider the evolution equation for the twist-3 quasipartonic
operators in double logarithmic approximation (DLA), which collects the
powers of the product $\alpha_S\ln Q^2/\mu^2 \ln 1/x$.
Quasipartonic operator form the closed set of high twist
operators allowing for the interpretation in terms of the parton model
\cite{BFKL}. They are responsible for the small $x_B$ asymptotics of
the structure function \cite{QP,VL}. The matrix
elements of quasipartonic operators depend only on the fraction $x_i$
of the partons momenta along the hadron momentum $p$ ($p^2 \simeq 0$).
The pure gluon channel will be studied below as dominating in the small
$x_B$ region. We shall take the quasipartonic operators, which in the axial
gauge $n_\mu A_\mu=0$ with a light-like vector $n$ dual to the hadron
momentum $p$, have a general form
$$
{\cal O}_{\mu_1, \mu_2, \mu_3}^{\,m_1,m_2,m_3}\,=\,
\Gamma_{\mu_1 \mu_1^\prime,\mu_2 \mu_2^\prime,\mu_3 \mu_3^\prime}^{\,abc}
\bigl((i\partial)^{m_1}A_{\mu_1^\prime}^a\bigr)
\bigl((i\partial)^{m_2}A_{\mu_2^\prime}^b\bigr)
\bigl((i\partial)^{m_3}A_{\mu_3^\prime}^c\bigr),
$$
where $m_i$ are positive integers, $\Gamma$ is a color and Lorentz tensor,
the particular form of which will not be important in what follows,
$\partial = n_\mu \partial_\mu$. The matrix element over a hadron state can
be expressed as
\begin{eqnarray}
\langle h|{\cal O}_{\mu_1, \mu_2, \mu_3}^{\,m_1,m_2,m_3}|h\rangle\,&=&\,
\int d\beta_1 d\beta_2 d\beta_3\,N_{\lambda_1 \lambda_2 \lambda_3}^{abc}
(\beta_1, \beta_2, \beta_3)
\nonumber \\
&\times&\,{\cal O}_{\lambda_1 \mu_1,\lambda_2 \mu_2,
\lambda_3 \mu_3}^{\,abc}\beta_1^{\,m_1}\beta_2^{\,m_2}\beta_3^{\,m_3},
\nonumber \\
{\cal O}_{\lambda_1 \mu_1,\lambda_2 \mu_2,\lambda_3 \mu_3}^{\,abc}\,&=&\,
\Gamma_{\mu_1 \mu_1^\prime,\mu_2 \mu_2^\prime,\mu_3 \mu_3^\prime}^{\,abc}
\varepsilon_{\mu_1^\prime}^{\lambda_1}\varepsilon_{\mu_2^\prime}^{\lambda_2}
\varepsilon_{\mu_3^\prime}^{\lambda_3}, \nonumber
\end{eqnarray}
where $\varepsilon_{\mu}^{\lambda}$ is the gluon polarization vector.
The parton correlation function
$N_{\lambda_1 \lambda_2 \lambda_3}^{abc}(x_1,x_2,x_3)$ has a meaning of
a hadron wavefunction integrated over partons transverse momenta,
the greatest transverse momentum being of the order $Q^2$.

The evolution equation generalizing twist-2 DGLAP equation is derived
in paper \cite{BFKL}. It has the form of $N$-particle one dimensional
Schr\"odinger-type equation with pairwise interaction between gluons,
\begin{equation}
\label{LLA}
Q^2\frac \partial {\partial Q^2 }
N_{\lambda_1,\ldots, \lambda_i,\lambda_j,\ldots,\lambda_N}
^{a_1,\ldots, a_i,a_j,\ldots,a_N}
(Q^2 ,x_1,\ldots ,x_N)=
\sum_{i<j}\int d\beta_i d\beta_j\,\delta (x_i+x_j-\beta_i-\beta_j)
\end{equation}
$$
\times \Phi_{\lambda_i \lambda_j,\lambda_i^\prime \lambda_j^\prime}
^{a_i a_j,a_i^\prime a_j^\prime}
(x_i,x_j; \beta_i,\beta_j)
N_{\lambda_1,\ldots, \lambda_i^\prime,\lambda_j^\prime,\ldots,
\lambda_N}^{a_1,\ldots, a_i^\prime,a_j^\prime,\ldots,a_N}
(Q^2 ,x_1,\ldots ,\beta_i,\ldots ,\beta_j,\ldots ,x_N)
$$
This equation sums up in the leading $\ln Q^2$ order the ladder-type
diagrams. For the twist $N$ case they comprise
the local operator vertex and $N$ gluons in
$t$-channel interacting through all possible $s$-channel gluons rungs.
The integrals in each ladder cell are ordered in LLA such that transverse
momentum in the above cell plays the role of ultraviolet cut-off for
the below one. The $Q^2$ value, being the greatest momentum
in the upper loop attached to the local operator vertex, is the ultraviolet
cut-off for the whole diagram. The evolution equation is obtained by taking
the derivative of the diagrams with respect to $\log Q^2$.
The kernel $\Phi_{\lambda_i \lambda_j,\lambda_i^\prime \lambda_j^\prime}
^{a_i a_j,a_i^\prime a_j^\prime} (x_i,x_j; \beta_i,\beta_j)$
is determined by the logarithmic part of the one-loop integral over
the transverse parton momentum $k_{\bot}$.

Generally the longitudinal momenta $x_i$ are not ordered in LLA, but
they have to be ordered in DLA to provide a large logarithm for each
ladder cell. The $x_i$ variables increase from the smallest values at
the local operator vertex to the order of unity ones in the lower part of a
diagram. In such a kinematics the logarithmic divergencies that occur in
every loop when $\beta_i\to 0$ is cut from below by
the longitudinal momentum in the upper cell. Thus DLA implies the loop
integrals in the evolution equation (\ref{LLA}) to be limited by
the condition
\begin{equation}
\label{order}
\beta_i,\beta_j \ll x_i,x_j,
\end{equation}
which means that the momenta below the $s$-channel rung ($x$) and above
it ($\beta$) are of different order of magnitude. The most
singular contribution comes in Eq.(\ref{LLA}) from the region where both
$\beta_i$ and $\beta_j$ tend to zero. The momentum
conservation allows it only if
\begin{equation}
\label{10}
x_i+x_j \ll \beta_i,\beta_j,
\end{equation}
that is $x_i\approx -x_j$ with logarithmic accuracy. Hereafter it
is convenient to assume the momenta directed upward to be positive, directed
downward to be negative.

In the logarithmic domain the kernel of the evolution equation in DLA can be
easily obtained by keeping the most singular in $\beta $ terms
in the gluon-gluon kernel,
\begin{equation}
\label{DLAker}
\Phi_{\lambda_1 ,\lambda_2 ,\lambda_1^\prime, \lambda_2^\prime}
^{a,b,c,d}(x_1,x_2;\beta_1,\beta _2)\,=\,2\delta_{\lambda_1 \lambda_2 }
\delta _{\lambda_1^\prime, \lambda_2^\prime}if^{acg}if^{bgd}\,
x_1\delta (x_1-x_2)\frac 1{\beta_1 \beta_2}.
\end{equation}
Here $\lambda_i,\lambda_i^\prime$ are the two dimensional transverse helicity
indices, $f^{abc}$ are the structure constants of the $SU(N_c)$ group. The
momenta $x_1,x_2$, are equal in DLA. They are positive but have
the opposite directions, one of them is incoming from below the other is
outgoing. The low-scale momenta $\beta_{1,2}$
are not supposed to be equal in DLA since the momentum transfer from below
$x_1-x_2$ is small only compared to the large momenta $x_{1,2}$
but is of the same order as the low-scale ones.

All $x_i$, $\beta_i$ momenta are supposed to be positive in the DLA kernel
(\ref{DLAker}), the sign is specified with an additional index
$\sigma = \{+,-\}$. Thus each gluon in the structure function is
characterized by the color index $a$, helicity $\lambda$, momentum value
$\beta$ and momentum direction $\sigma$. The interaction occurs only between
the gluons with the opposite $\sigma$.

There are two possible color structures for twist-3 operators -
the odderon-like $d^{abc}$ and gluon-like $f^{abc}$. Both of them go through
the equation resulting into $N_c/2$ factor. This simplifies the color
structure of the correlation function,
$N_{\lambda_1,\lambda_2,\lambda_3}^{abc}=
d^{abc} F_{\lambda_1,\lambda_2,\lambda_3}$ or
$N_{\lambda_1,\lambda_2,\lambda_3}^{abc}=f^{abc}
F_{\lambda_1,\lambda_2,\lambda_3}$, and the action of the kernel can be
written as
\begin{equation}
\label{H12}
H_{12} F_{\lambda_1,\sigma_1,\lambda_2,\sigma_2,\lambda_3,\sigma_3}
(x_1,x_2,x_3)\,=\,\frac 14\,\overline\alpha\,\delta_{\sigma_1,-\sigma_2}\,
\delta_{\lambda_1,\lambda_2}\delta_{\lambda_1^\prime,\lambda_2^\prime}\,
x_1\,\delta(x_1-x_2)
\end{equation}
$$
\times \int_0^{x_1}\frac{d\beta_1}{\beta_1}
\int_0^{x_2}\frac{d\beta_2}{\beta_2}\,
\bigl[\,F_{\lambda_1^\prime,\sigma_1,\lambda_2^\prime,\sigma_2,
\lambda_3,\sigma_3}(\beta_1,\beta_2,x_3)\,\pm\,
F_{\lambda_2^\prime,\sigma_2,\lambda_1^\prime,\sigma_1,
\lambda_3,\sigma_3}(\beta_2,\beta_1,x_3)\,\bigr],
$$
$$
\overline\alpha \,\equiv \, N_c\,\frac{\alpha_S}{\pi},
$$
and similarly for $H_{23}$, $H_{13}$. Two terms in the r.h.s. (\ref{H12})
represent the sum of the $s$- and $u$-channel diagrams (only even momentum
operators survive in this sum for the twist-2 case).

\noindent
3. Instead of direct solving of evolution equation we adopt here an other
approach similar to reggeon calculus and more suitable to find the asymptotics
of the structure function. To this end we rewrite the formal solution of
the evolution equation with a given initial condition $F_0$,
$$
F(Q^2)\,=\,e^{H\log Q^2/\mu^2}F_0,
$$
through the Mellin transform as
$$
F(Q^2)\,=\,\int \frac{d\nu}{2\pi i}\,\left(\frac{Q^2}{\mu^2}\right)^\nu
\frac 1\nu\,\frac 1{1-\frac 1\nu H}\,F_0,
$$
where the integral runs along the imaginary axis to the right from all
singularities. The equation
\begin{equation}
\label{Fnu}
F(\nu)\,=\,F_0\,+\,\frac 1\nu H\,F(\nu)
\end{equation}
can be treated as the Bethe-Solpeter equation in the theory described by
effective action
\begin{eqnarray}
S\,&=&\,\int dx\,\Phi^*_{\lambda \sigma}(x)\,\Phi_{\lambda \sigma}(x)
\nonumber \\
&+&\,
\frac 14\,\frac{\overline \alpha}{\nu}\,\int dx\,x\,
\frac{d\beta_1}{\beta_1}
\,\frac{d\beta_2}{\beta_2}\,\Phi^*_{\lambda_1 \sigma_1}(x)\,
\theta(x-\beta_1)\,\Phi_{\lambda_2 \sigma_2}(\beta_1)
\nonumber \\
&\times &\,\delta_{\lambda_1 \lambda_2}\delta_{\lambda_3 \lambda_4}
\bigl[\delta_{\sigma_1 \sigma_2}\delta_{\sigma_3 \sigma_4} \pm
\delta_{\sigma_1 \sigma_4}\delta_{\sigma_2 \sigma_3}\bigr]
\,\delta_{\sigma_2, -\sigma_4}\,
\Phi^*_{\lambda_3 \sigma_3}(x)\,
\theta(x-\beta_2)\,\Phi_{\lambda_4 \sigma_4}(\beta_2).\nonumber
\end{eqnarray}
The solution to eq.(\ref{Fnu}) is given by the convolution with the Green
function
\begin{equation}
\label{gf}
F_{\lambda_i^\prime \sigma_i^\prime}(\nu,\beta_i)\,=\,\int dx_i\,
\varphi_{\lambda_i \sigma_i}(x_i)\,
G_{\lambda_i \sigma_i ,\, \,\lambda_i^\prime \sigma_i^\prime}
(\nu;x_i,\beta_i)
\end{equation}
calculated in the effective theory,
$$
G_{\lambda_i \sigma_i, \lambda_i^\prime \sigma_i^\prime}
(\nu; x_i,\beta_i)\,=\,
\prod_{i=1}^3\langle \Phi_{\lambda_i^\prime \sigma_i^\prime}(x_i)\,
\Phi^*_{\lambda_i \sigma_i}(\beta_i)\rangle.
$$
The initial hadron wavefunction
$\varphi_{\lambda_i \sigma_i}(x_i)$ is to be taken at the low
$Q^2$ scale. It can't be find perturbatively, but its precise form do not
influence large $Q^2$ asymptotics. For definiteness we shall consider below
the moments, that is we take
\begin{equation}
\label{xmom}
\varphi(x_i)=\prod_{i=1}^3 x_i^{n_i}
\end{equation}
with the integers $n_i \ge 0$.
\begin{center}
\global\newsavebox{\RUG}
\savebox{\RUG}(0,0)
{
\begin{picture}(3500,500)
\drawline\fermion[\E\REG](500,250)[2500]
\put(\pfrontx,\pfronty){\circle*{200}}
\put(\pbackx,\pbacky){\circle*{200}}
\end{picture}
}
\global\newsavebox{\DOT}
\savebox{\DOT}(0,0)
{
\begin{picture}(3500,500)
\startphantom
\drawline\fermion[\E\REG](500,250)[2500]
\stopphantom
\put(\pmidx,\pfronty){\circle*{300}}
\end{picture}
}
\begin{picture}(10000,10000)
\drawline\fermion[\N\REG](500,500)[8000]
\drawline\fermion[\N\REG](3000,500)[8000]
\global \advance \pfrontx by -4500
\global \Yone = \pmidy
\put(\pfrontx,\Yone){$\sum $}
\startphantom
\drawline\fermion[\E\REG](\pbackx,\pbacky)[4000]
\stopphantom
\put(\pmidx,\Yone){$=$}
\drawline\fermion[\SE\REG](\pbackx,\pbacky)[1500]
\drawline\gluon[\S\REG](\pbackx,\pbacky)[5]
\drawline\fermion[\NE\REG](\gluonfrontx,\gluonfronty)[1500]
\drawline\fermion[\SW\REG](\gluonbackx,\gluonbacky)[1500]
\drawline\fermion[\SE\REG](\gluonbackx,\gluonbacky)[1500]
\drawoldpic\RUG(1750,1500)
\drawoldpic\DOT(1750,2500)
\drawoldpic\DOT(1750,3500)
\drawoldpic\DOT(1750,4500)
\drawoldpic\RUG(1750,5500)
\drawoldpic\RUG(1750,6500)
\drawoldpic\RUG(1750,7500)
\end{picture}
\vskip 0.5 cm
Figure 1
\end{center}

We shall find the Green function by iterating
the Bethe-Solpeter equation. We start from the two gluon ladder, or "reggeon",
which will be a main building block in the further
proceeding. It is schematically shown in Fig.1, where the solid lines denote
gluon. Iterations of the two-particle kernel results into the matrix
\begin{eqnarray}
\label{gR}
\widehat g_{\lambda_1\,\lambda_2\,,\,\lambda_1^\prime\,\lambda_2^\prime}
(x_1,x_2;\beta_1,\beta_2)
\,&=&\,\frac 12\,\delta_{\lambda_1 \lambda_2}
\delta_{\lambda_1^\prime \lambda_2^\prime}\,
\frac 1{\beta_1}\frac 1{\beta_2}\,\widehat g_{12}(\nu; x_1,\{\beta_1,\beta_2\})\,\\
&\times&\,x_1\,\delta(x_1-x_2). \nonumber
\end{eqnarray}
Here $\{\beta_1,\beta_2\}\equiv max\{\beta_1,\beta_2\}$ and matrix $g_{ik}$
acts on the sign variables $\sigma_i$, $\sigma_k$ as
$$
\widehat g_{ik}(\nu;x,\beta)\,\equiv \,
g_{\sigma_i\sigma_k,\sigma_i^\prime\sigma_k^\prime}(\nu;x,\beta)\,=\,
\left(\frac{I \pm P_{ik}}{2}A_{ik}\right)_{\sigma_i\sigma_k,
\sigma_i^\prime\sigma_k^\prime}\,g(\nu;x,\beta),
$$
where operator $P_{ik}$ permutes the indices $\sigma_i$, $\sigma_k$,
matrix $A_{ik}$ permits the interaction only between the partons with
opposite momentum signs,
$$
\bigl(A_{ik}\bigr)_{\sigma_i\sigma_k,\,\sigma_i^\prime\sigma_k^\prime}\,=\,
\delta_{\sigma_i \sigma_i^\prime}\,\delta_{\sigma_k \sigma_k^\prime}\,
\delta_{\sigma_i^\prime, -\sigma_k^\prime},
$$
and
\begin{equation}
\label{gnuj}
g(\nu;x,\beta)\,=\,
\int \frac{dj}{2\pi i}\,\left(\frac\beta x \right)^{-j}
\frac {\overline\alpha}{2\nu j-\overline\alpha}.
\end{equation}

\noindent
4. The expression (\ref{gR}) leads to the usual structure function in
the twist-2 case. Indeed, the general form of twist-2 spin $J$ gluon
operator ($F_1$ structure function) convoluted with the gauge fixing
vector $n$ is
$$
n_{\mu_1}\cdots n_{\mu_J}{\cal O}_{\mu_1,\ldots, \mu_J}\,=\,
\bigl(i\partial \bigr)A_\nu\,\bigl(i\partial \bigr)^{J-1}A_\nu.
$$
It results into the vertex
$$
\delta_{\lambda_1 \lambda_2}\beta_1^{m_1}\beta_2^{m_2}\delta(\beta_1+\beta_2)
$$
with $m_1=1$, $m_2=J-1$. There is no momentum transfer through the operator,
this is the reason for the momentum delta-function. This vertex
should be integrated with the ladder function (where $\overline \alpha/2$
is replaced with $\overline\alpha$ for the color singlet).
The integration has to be done with
account of both signs of the $\beta_{1,2}$ variables, which implies the sum
over $(+ -)$ and $(- +)$ initial sign states. For the $(+ -)$ final state,
that is for the positive $x_1$, we get
$$
M_2(J)\,=\,-\bigl[1+(-1)^J\bigr]\,\delta_{\lambda_1 \lambda_2}
x_1^J\delta(x_1-x_2)\,\int\frac{dj}{2\pi i}\,
\frac {\overline\alpha}{\nu(J-1)-\overline\alpha}
$$
and the same for the $(- +)$ state. Thus we have reproduced the double
logarithmic twist-2 anomalous dimension
$$
\gamma_2(J)\,=\,\frac {\overline\alpha}{J-1}
$$
together with the selection rule allowing only for the even $J$ values.

\noindent
5. We consider the diagrams for the Green function occurring in the eq.(\ref{gf}).
The general sum of the three-gluon ladder diagrams can be equivalently
presented as a sum of two-gluon ladders ("reggeons") developing between each
gluon pairs accompanied with a third single gluon as is shown in Fig.2 -- 4.
By employing this representation all diagrams can be summed up in a closed
form. The Green function reeds
\begin{equation}
\label{gfd}
G_{d,f}^{tot}\,=\,\sum_{\{i\},\{i^\prime\}}
\overline
G_{d,f}(x_{i},\sigma_{i},\lambda_{i}\,|\,
\beta_{i^\prime},\sigma_{i^\prime},\lambda_{i^\prime})
\end{equation}
where the sum is taken over independent permutations of the incoming and
outgoing particles while the functions
$\overline G_{d,f}$ stand for
the diagrams with a fixed order of the external lines.
The symbols $d$ and $f$ label the Green functions for the $d^{abc}$ and
$f^{abc}$ color structures. The formula (\ref{gfd}) implies the simultaneous
permutations of all quantum numbers, that is momenta, helicities and colors,
which means symmetrization with respect $\{x_i,\sigma_i,\lambda_i \}$
(or $\{\beta_j,\sigma_j^\prime,\lambda_j^\prime \}$) pair for $d^{abc}$
tensor and antisymmetrization for $f^{abc}$ tensor.
\begin{center}
\begin{picture}(1000,2000)
\drawline\fermion[\SE\REG](0,0)[1500]
\global\advance \fermionfrontx by -5000
\put(\fermionfrontx,\fermionfronty){$\beta_1,\sigma_1^\prime,
\lambda_1^\prime$}
\drawline\gluon[\S\REG](\pbackx,\pbacky)[3]
\global\advance \pmidx by 500
\put(\pmidx,\pmidy){$(12)$}
\drawline\fermion[\NE\REG](\gluonfrontx,\gluonfronty)[1500]
\global\advance \fermionbackx by 500
\put(\fermionbackx,\fermionbacky){$\beta_2,\sigma_2^\prime,\lambda_2^\prime$}
\drawline\fermion[\SW\REG](\gluonbackx,\gluonbacky)[1500]
\global\advance \fermionbackx by -5000
\put(\fermionbackx,\fermionbacky){$x_1,\sigma_1,\lambda_1$}
\drawline\fermion[\SE\REG](\gluonbackx,\gluonbacky)[1500]
\startphantom
\drawline\fermion[\E\REG](\pbackx,\pbacky)[6500]
\stopphantom
\global\advance \fermionfrontx by 500
\put(\fermionfrontx,\fermionfronty){$x_2,\sigma_2,\lambda_2$}
\drawline\fermion[\N\REG](\pbackx,\pbacky)[5300]
\global\advance \fermionfrontx by 500
\put(\fermionfrontx,\fermionfronty){$x_3,\sigma_3,\lambda_3$}
\global\advance \fermionbackx by 500
\put(\fermionbackx,\fermionbacky){$\beta_3,\sigma_3^\prime,\lambda_3^\prime$}
\end{picture}
\vskip 2.5 cm
Figure 2
\end{center}

The effective diagrams constructed from the two-gluon ladder and
gluon line turns out to be rather simple to calculate the Green
function by the direct summation. The result is presented by the sum
of three contributions. Two of them are degenerate in the sense that they
comes from the finite number of iterations.
The first one includes the "reggeon" only once. There are 3 diagrams of this
type for 3 various gluon pairs combined into the ladder. One of them for
the pair $12$ is shown in Fig.2,
\begin{eqnarray}
\overline G_{d,f}^{(I)}\,(x_i,\beta_i)_{(12)}\,&=&\,\frac 12\,
\delta_{\lambda_1 \lambda_2}\delta_{\lambda_3
\lambda_3^\prime} \delta_{\lambda_1^\prime \lambda_2^\prime}\,
\frac 1{\beta_1 \beta_2}\, \widehat g_{12}(\nu;x_1,\{\beta_1, \beta_2\})
\nonumber \\
&\times &\, x_1\delta(x_1-x_2)\delta(x_3-\beta_3)\,
\delta_{\sigma_3 \sigma_3^\prime}. \nonumber
\end{eqnarray}
The others can be obtained by permutations of the indices $(123) \to (231)$
and $(123) \to (132)$.
\begin{center}
\begin{picture}(1000,2000)
\drawline\fermion[\SE\REG](0,0)[1500]
\global\advance \fermionfrontx by -5000
\put(\fermionfrontx,\fermionfronty){$\beta_1,\sigma_1^\prime,
\lambda_1^\prime$}
\drawline\gluon[\S\REG](\pbackx,\pbacky)[3]
\global\advance \pmidx by 500
\put(\pmidx,\pmidy){$(12)$}
\drawline\fermion[\NE\REG](\gluonfrontx,\gluonfronty)[1500]
\global\advance \fermionbackx by 500
\put(\fermionbackx,\fermionbacky){$\beta_2,\sigma_2^\prime,\lambda_2^\prime$}
\drawline\fermion[\SW\REG](\gluonbackx,\gluonbacky)[4000]
\global\advance \fermionbackx by -5000
\put(\fermionbackx,\fermionbacky){$x_1,\sigma_1,\lambda_1$}
\drawline\fermion[\SE\REG](\gluonbackx,\gluonbacky)[2500]
\drawline\gluon[\S\REG](\pbackx,\pbacky)[3]
\global\advance \pmidx by 500
\put(\pmidx,\pmidy){$(23)$}
\drawline\fermion[\NE\REG](\gluonfrontx,\gluonfronty)[4000]
\global\advance \fermionbackx by 500
\put(\fermionbackx,\fermionbacky){$\beta_3,\sigma_3^\prime,\lambda_3^\prime$}
\drawline\fermion[\SW\REG](\gluonbackx,\gluonbacky)[1500]
\global\advance \fermionbackx by -5000
\put(\fermionbackx,\fermionbacky){$x_2,\sigma_2,\lambda_2$}
\drawline\fermion[\SE\REG](\gluonbackx,\gluonbacky)[1500]
\global\advance \fermionbackx by 500
\put(\fermionbackx,\fermionbacky){$x_3,\sigma_3,\lambda_3$}
\end{picture}
\vskip 4 cm
Figure 3
\end{center}

The second contribution arises from the diagram with the two "reggeons".
Fig.3 presents the diagram where the gluon pair $12$ switches to the pair
$23$,
\begin{eqnarray}
\label{II}
\overline G_{d,f}^{(I\!I)}\,(x_i,\beta_i)_{(23)(12)}\,&=&\,\frac 14\,
\delta_{\lambda_2 \lambda_3}\delta_{\lambda_1\lambda_3^\prime }
\delta_{\lambda_1^\prime \lambda_2^\prime}\,
\frac 1{\beta_1 \beta_2 \beta_3}\,
\widehat g_{\,23}(\nu;x_1,\{\beta_1, \beta_2\})  \\
&\times&\,
\widehat g_{12}(\nu;x_2,\{x_1, \beta_3\})\,x_2\delta(x_2-x_3). \nonumber
\end{eqnarray}
The other 5 terms result in this case from (\ref{II}) after independent
permutations $(123) \to (231)$ and $(123) \to (132)$ of the upper and lower
(in the sense of Fig.3) indices but excluding the equal ones. In other words
the sum is taken over various ways to combine the incoming and outgoing
gluons into different two-particle ladders.

The contributions starting with the three "reggeons" develop a regular
series which can be written as
\begin{eqnarray}
\label{III}
\overline G_{d,f}^{(I\!I\!I)}(x_i,\beta_i)_{(12)(12)}\,&=&\,\frac 14\,
\delta_{\lambda_1 \lambda_2}\delta_{\lambda_3 \lambda_3^\prime}
\delta_{\lambda_1^\prime \lambda_2^\prime} \,
\frac 1{\beta_1 \beta_2 \beta_3}\,
\widehat g_{12}(\nu;x_1,x_3)
 \\
&\times&\,\int \frac{d\beta}{\beta}\,W_{d,f}(x_3,\nu;\{\beta_3,\beta \})\,
\widehat g_{12}(\nu;\beta,\{\beta_1, \beta_2\}),
\nonumber \\
\label{Wbeta}
W_{d,f}(\nu;x,\beta)\,&=&\,\frac 12\,\widehat g(\nu; x,\beta) \\
&+&\,
\frac 14\,\int \frac{d\beta^\prime}{\beta^\prime}\,
\widehat g(\nu;x,\beta^\prime)\,
\widehat g(\nu;\beta^\prime,\beta)\,+\,\cdots . \nonumber
\end{eqnarray}
Fig.4 shows the diagram corresponding either to the first term in
eq.(\ref{Wbeta}) or to all terms in (\ref{Wbeta}) if the middle "reggeon" is
replaced with the full function $W_{d,f}$.
\begin{center}
\begin{picture}(1000,2000)
\drawline\fermion[\SE\REG](0,0)[1500]
\global\advance \fermionfrontx by -5000
\put(\fermionfrontx,\fermionfronty){$\beta_1,\sigma_1^\prime,
\lambda_1^\prime$}
\drawline\gluon[\S\REG](\pbackx,\pbacky)[2]
\global\advance \pmidx by 500
\put(\pmidx,\pmidy){$(12)$}
\global \Yone = \gluonbacky
\global \negate \Yone
\drawline\fermion[\NE\REG](\gluonfrontx,\gluonfronty)[1500]
\global\advance \fermionbackx by 500
\put(\fermionbackx,\fermionbacky){$\beta_2,\sigma_2^\prime,\lambda_2^\prime$}
\drawline\fermion[\SE\REG](\gluonbackx,\gluonbacky)[4000]
\drawline\gluon[\S\REG](\fermionbackx,\fermionbacky)[2]
\drawline\fermion[\SE\REG](\gluonbackx,\gluonbacky)[5000]
\global\advance \fermionbackx by 500
\put(\fermionbackx,\fermionbacky){$x_3,\sigma_3,\lambda_3$}
\drawline\fermion[\NE\REG](\gluonfrontx,\gluonfronty)[5000]
\global\advance \fermionbackx by 500
\put(\fermionbackx,\fermionbacky){$\beta_3,\sigma_3^\prime,\lambda_3^\prime$}
\drawline\fermion[\SW\REG](\gluonbackx,\gluonbacky)[4000]
\global \advance \Yone by \pbacky
\drawline\gluon[\S\REG](\fermionbackx,\fermionbacky)[2]
\global\advance \pmidx by 500
\put(\pmidx,\pmidy){$(12)$}
\drawline\fermion[\N\REG](\fermionbackx,\fermionbacky)[\Yone]
\drawline\fermion[\SW\REG](\gluonbackx,\gluonbacky)[1500]
\global\advance \fermionbackx by -5000
\put(\fermionbackx,\fermionbacky){$x_1,\sigma_1,\lambda_1$}
\drawline\fermion[\SE\REG](\gluonbackx,\gluonbacky)[1500]
\global\advance \fermionbackx by 500
\put(\fermionbackx,\fermionbacky){$x_2,\sigma_2,\lambda_2$}
\end{picture}
\vskip 6 cm
Figure 4
\end{center}
There are 8 other terms besides (\ref{III}), which can be obtained from it
by independent permutations $(123) \to (231)$ and $(123) \to (132)$
of the upper and lower indices in the Fig.4.

The Mellin transform (\ref{gnuj}) turns the convolutions over
$\beta^\prime$ into the usual products, making the series (\ref{Wbeta})
to be equivalent to the purely matrix problem,
\begin{equation}
\label{W}
W\,=\,\frac 12\frac{\overline\alpha}{\nu j}H\,+\,
\bigl(\frac 12 \frac{\overline\alpha}{\nu j}H\bigr)^2\,+\,
\cdots \,=\,H\,\frac 1{2\frac{\nu j}{\alpha} - H},
\end{equation}
where the "reggeons" "dissociate" into two-particle interaction,
$$
H\,=\,\sum H_{i,i+1},\qquad H_{i,k}\,=\,\frac 12\,\delta_{\lambda_i \lambda_k}
\delta_{\lambda_i^\prime \lambda_k^\prime}\,\frac{1\pm P_{ik}}{2}\,A_{ik},
$$
and the problem is reduced to the inversion of the finite matrix.

The above iterations exhibit that at each step two momenta with opposite
directions have the same value much larger than the value of the third
momentum ($x_{2,\,3}\gg x_1$ in Fig.3 and $x_{1,2}\gg x_3$ in Fig.4).
This property expreses the longitudinal momentum conservation within
the DLA accuracy -- the sum of all momenta is small compared to their
natural scale.

\noindent
6. The Green function is convoluted over variables $\beta_i$ and helicities
indices $\lambda_i^\prime$ with the operator vertex given by the expression
\begin{equation}
\label{O}
O_{\lambda_1^\prime,\mu_1, \lambda_2^\prime,\mu_2, \lambda_3^\prime,\mu_3}
(\beta_{\,1},\beta_{\,2},\beta_{\,3})
\,=\,
\Gamma_{\lambda_1^\prime,\mu_1, \lambda_2^\prime,\mu_2, \lambda_3^\prime,
\mu_3}\,\beta_{\,1}^{\,m_1}\beta_{\,2}^{\,m_2} \beta_{\,3}^{\,m_3}
\delta(\beta_{\,1}+\beta_{\,2}+\beta_{\,3}),
\end{equation}
where the longitudinal $\delta$-function corresponds to forward kinematics
without momentum transfer. Taking then the moments with respect variables
$x_i$ (\ref{xmom}) we get the Green function in the moments representation,
$G_{d,f}^{tot}(m_i,n_i)$. Note, that the integrals over $x_i$, $\beta_i$
imply the positive as well as the negative values of the momenta.
The negative values are described through the sign variables
$\sigma_i = \pm$, for example, the configuration where $\beta_1<0$,
$\beta_{\,2,\,3}>0$ is associated with the state $(-,+,+)$ and similarly
for $x_i$. The integral over all sign configurations is recovered by the sum
over all initial $\sigma_i^\prime$ and final $\sigma_i$ values. As a result
the Green function written in terms of the moments takes into account both
signs of $\beta_i$ and $x_i$ and does not contain the auxiliary variables
$\sigma_i$,
\begin{equation}
G_{d,f}^{tot}(m_{i^\prime},n_i)\,=\,\sum_{\{i^\prime\},\{i\}}
\delta_{\lambda_{i_1}\lambda_{i_2}}
\delta_{\lambda_{i_1^\prime}\lambda_{i_2^\prime}}
\delta_{\lambda_{i_3}\lambda_{i_3^\prime}}
\overline G_{d,f}
(m_{i_1^\prime},m_{i_2^\prime},m_{i_3^\prime};n_{i_1},n_{i_2},n_{i_3}).
\end{equation}
The sum here means the independent symmetrization with respect the pairs
$\{m_{i^\prime},\lambda_{i^\prime}\}$ and $\{n_i,\lambda_i\}$ for $d^{abc}$
and antisymmetrization for $f^{abc}$ structures. The helicities
$\lambda_i^\prime$ should be convoluted with the tensor
$\Gamma_{\lambda_i^\prime,\mu_i}$ specifying the operator vertex (\ref{O}).

Separating the common factors the functions $\overline G_{d,f}(m_i,n_i)$
takes the form
\begin{eqnarray}
\overline G_{d,f}(m_i,n_i)\,&=&\,\bigl[1+(-1)^{m+n}\bigr]\,
\bigl((-1)^{n_1}\pm (-1)^{n_2}\bigl)\,
\biggl[\frac{(-1)^{m_1}}{m_2}\pm\frac{(-1)^{m_2}}{m_1}\biggr]\, \nonumber \\
&\times& \,\frac 1{m+n+2}\,G_{d,f}(m_i,n_i),
\end{eqnarray}
$$
m\,\equiv\,m_1\,+\,m_2\,+\,m_3, \qquad n\,\equiv\,n_1\,+\,n_2\,+\,n_3,
$$
where $+$ and $-$ stand for $d$ and $f$ structures, respectively, and
the function $G_{d,f}$ is expressed in terms of the three
contributions considered above,
$$
G_{d,f}\,=\,G_{d,f}^{(I)}\,+\,G_{d,f}^{(I\!I)}\,+\,G_{d,f}^{(I\!I\!I)}.
$$
The first contribution yields
\begin{equation}
\label{GI}
G_{d,f}^{(I)}(m_i,n_i)\,=\,\frac 34\,
\frac {\overline\alpha}{2\nu j-\overline\alpha},
\end{equation}
$$
j\,=\,m+n_3,
$$
while the second one takes the form:
\begin{eqnarray}
\label{GIId}
G_{d}^{(I\!I)}(m_i,n_i)\,&=&\,\frac 38\,
\frac {\overline\alpha}{2\nu (m-1)-\overline\alpha}\,
\frac {\overline\alpha}{2\nu j-\overline\alpha} \\
\label{GIIf}
G_{f}^{(I\!I)}\,&=&\,0.
\end{eqnarray}
($G_{f}^{(I\!I)}$ vanishes after antisymetrization over end points).
The third contribution with the matrix (\ref{W}) inverted
reads
\begin{eqnarray}
\label{GIIId}
G_{d}^{(I\!I\!I)}\,&=&\,\frac 3{8}\,
\frac {\overline\alpha}{2\nu (m-1)-\overline\alpha}\,
\frac {\overline\alpha}{2\nu j-\overline\alpha}\,
\frac{\overline\alpha}{4\,\nu (m-1)-3\,\overline\alpha},\\
\label{GIIIf}
G_{f}^{(I\!I\!I)}\,&=&\,\frac 3{8}\,
\frac {\overline\alpha}{2\nu (m-1)-\overline\alpha}\,
\frac {\overline\alpha}{2\nu j-\overline\alpha}\,
\frac{\overline\alpha}{4\,\nu (m-1) -\overline \alpha},
\end{eqnarray}

\noindent
7. The asymptotic behavior of the structure function for Bjorken variable
$x_B \to 0$ is determined by the rightmost singularity in the variable $J$,
which has the meaning of the local operator spin continued to
the complex plane. The spin of quasipartonic operator $J=m$, therefore
one needs to continue the function $\overline G_{d,f}(m_i,n_i)$
to $m \to 1$ formally keeping the other variables $m_i$, $n_i$ to be
fixed. Because of the signature-like factors $(-1)^{m+n}$
the terms with even or odd $m$ are to be treated separately.
Note that in a general LLA case this continuation is non trivial since
the mixing matrix describing the evolution has a rank depending
on $J$ \cite{Br}.
The explicit form of the DLA solutions (\ref{GI}) - (\ref{GIIIf}) makes
the continuation much more simple and straightforward.
The obtained results show that
the anomalous dimension for $d^{abc}$ color structure is
\begin{equation}
\label{gammad}
\gamma_d(J)\,=\,\frac 34\,\frac{\overline\alpha}{J-1}.
\end{equation}
The main singularity for $f^{abc}$ structure is actually given by the pole
of 2 gluon state,
\begin{equation}
\label{gammaf}
\gamma_f(J)\,=\,\frac 12\frac{\overline\alpha}{J-1}.
\end{equation}
The contribution of "developed" 3 gluon ladder is more weak in this channel,
$\gamma_f^\prime(J) = \gamma_f(J)/2$. The contributions of other
singularities are strictly speaking beyond the DLA accuracy since they
produce the extra positive powers of $x_B$.

The DLA anomalous dimension (\ref{gammad}) is smaller compared to those,
which can be derived from the direct solution of BFKL equation obtained in
refs.\cite{VL,WJ,BLV,Kor}. The possible reason for this is that the known
BFKL solutions founded for Odderon do not really correspond to quasipartonic
operators of twist 3. In this case the DLA result (\ref{gammad}) could
indicate to an existence of another solution of quasipartonic type.

\smallskip
\noindent
{\bf Acknowledgments:} The author is grateful to L.N.Lipatov
for helpful discussions.\\
\smallskip
This work is supported by grant RFBR 01-02-17095.

\end{document}